\pdfoutput=1

\documentclass[10pt,pra,twocolumn,superscriptaddress,showpacs]{revtex4}

\usepackage{graphicx}
\usepackage{amssymb}
\usepackage{times}
\usepackage{amsmath}

\newcommand{\bra}[1]{\langle#1|}
\newcommand{\ket}[1]{|#1\rangle}

\begin{document}

\title{Distinguishability of Gaussian States in Quantum Cryptography using Post-Selection}

\author{Christian~Weedbrook}\email{christian.weedbrook@gmail.com} \affiliation{Department of Physics,
University of Queensland, St Lucia, Queensland 4072, Australia}

\author{Daniel J. Alton}
\altaffiliation{Current address: Norman Bridge Laboratory of Physics 12-33, California Institute of Technology, Pasadena, California 91125, USA.} \affiliation{Quantum Optics Group, Department
of Physics, Faculty of Science, Australian National University,
ACT 0200, Australia}

\author{Thomas~Symul} \affiliation{Quantum Optics Group, Department
of Physics, Faculty of Science, Australian National University,
ACT 0200, Australia}

\author{Ping~Koy~Lam}\affiliation{Quantum Optics Group, Department
of Physics, Faculty of Science, Australian National University,
ACT 0200, Australia}

\author{Timothy~C.~Ralph} \affiliation{Department of Physics, University of Queensland, St Lucia, Queensland 4072, Australia}

\date{\today}

\begin{abstract}

We consider the distinguishability of Gaussian states from the view point of continuous-variable quantum cryptography using post-selection. Specifically, we use the probability of error to distinguish between two pure coherent (squeezed) states and two particular mixed symmetric coherent (squeezed) states where each mixed state is an incoherent mixture of two pure coherent (squeezed) states with equal and opposite displacements in the conjugate quadrature. We show that the two mixed symmetric Gaussian states (where the various components have the same real part) never give an eavesdropper more information than the two pure Gaussian states. Furthermore, when considering the distinguishability of squeezed states, we show that varying the amount of squeezing leads to a ``squeezing" and ``anti-squeezing" of the net information rates. 

\end{abstract}


\maketitle

\section{Introduction}

The laws of quantum mechanics tell us that in general it is impossible to
perfectly distinguish between two non-orthogonal quantum states
\cite{Nielsen00a}. This limitation imposed by quantum measurement
theory \cite{Helstrom76a} is inherent in a number of continuous-variable (CV) quantum information \cite{S.L.Braunstein2005} applications, including quantum cloning and the security
of quantum cryptography protocols (e.g., see \cite{Cerf2007}). Closely related to this is quantum state discrimination \cite{A.Chefles2000,Bergou2004} which is concerned with the distinguishability of quantum states. There are two commonly used distinguishability techniques \cite{A.Chefles2000,Bergou2004}: (1) minimum error discrimination and (2) unambiguous state discrimination. In minimum error discrimination, a number of approaches have been developed where quantum states can be distinguished provided we allow a certain amount of uncertainty or error in the measurement results. On the other hand, unambiguous state discrimination is an error free discrimination process but relies on the fact that sometimes the observer gets an inconclusive result.

Previous work on the distinguishability of CV quantum states includes: calculating the Bures distance between two (displaced) squeezed thermal states \cite{J.Twamley1996,Gh.-S.Paraoanu1998}, unambiguous discrimination of symmetric coherent states \cite{A.Chefles1998} using linear optics \cite{Elk2002}, binary optical communication for single and entangled modes in realistic channels \cite{Oli04}, distinguishing single-mode Gaussian states using homodyne detection \cite{H.Nha2005}, coherent state estimation with minimal disturbance \cite{Andersen2006}, using the quantum Chernoff bound as the distinguishability measure \cite{Calsamiglia2008} and computable bounds for Gaussian state discrimination \cite{S.Pirandola2008b}. Furthermore, various techniques for optimally distinguishing pure optical coherent states with minimum error have been investigated theoretically \cite{Dol73,Ken73,Tak05,Sas96,Tak08,Ger04} and also experimentally \cite{Coo07,Wit08}.

In this paper, we consider a specific distinguishability situation in terms of the CV distinguishability of Gaussian states (in particular coherent and squeezed states) from the view point of CV quantum key distribution (CV-QKD) \cite{F.Grosshans2003,F.Grosshans2002,Silberhorn2002,C.Weedbrook2004,S.Pirandola2008}. The security of CV-QKD is fundamentally based on the inability of an eavesdropper to perfectly distinguish between
non-orthogonal quantum states \cite{Nielsen00a}. Here we look
at how much information a potential eavesdropper can gain when
trying to distinguish between two pure coherent states as opposed
to distinguishing between two mixed coherent states where each mixed state is an incoherent mixture of two pure coherent states with equal and opposite displacements in the conjugate quadrature. This is of
particular interest in CV-QKD schemes which use the original post-selection protocol
\cite{Silberhorn2002}, where it is often accepted that an
eavesdropper's knowledge can be upper bounded by assuming that she obtains more information in the case of
distinguishing between two pure coherent states than two mixed coherent states of equal phase-space separation \cite{Silberhorn2002,M.Heid2007,Lance2005,T.Symul2007}. It may have been anticipated that, given our particular distinguishability configuration in phase space, two mixed coherent states might be more distinguishable than two pure coherent states. However, this is not the case, and consequently, we show that this assumption in post-selection based CV-QKD is valid. In addition, we extend the coherent state case to include the distinguishability of squeezed states. We show that a ``squeezing" and ``anti-squeezing" of the net information rates occurs when varying the amount of squeezing. Furthermore, we see the effect (for both coherent and squeezed states) that after a certain amount of phase-space separation the two mixed Gaussian states start ``behaving" like the two pure Gaussian states in that the amount of information in distinguishing them both is equal.
We also briefly compare the probability of errors from using an optimal POVM (which corresponds to our distinguishability measure) to the more practical, and commonly used, quadrature projective measurement. 

This paper is structured as follows. In Section II we introduce the probability of error as our measure of distinguishability. Sections III and IV analyze the distinguishability of pure and mixed coherent and squeezed states, respectively. Finally, Section V offers a discussion with concluding remarks.

\section{Distinguishability Measure}

In this section we introduce our measure of distinguishability of CV
quantum states: the probability of error $p_e$. We
point out that there are other quantum distinguishability measures
including the Kolmogorov distance, the Bhattacharyya
coefficient and the Shannon distinguishability
(for a review of these measures, see e.g., Fuchs and van de Graaf
\cite{Fuchs99a}).

\subsection{Probability of Error}

A benefit of the probability of error is that it is related
to the trace norm distance, or simply the trace distance $D$,
between the two density matrices of the states being distinguished
and hence can be readily calculated. Furthermore, the corresponding
Shannon information can be determined directly from the
probability of error measure as we will soon see. It was originally shown by Helstrom
\cite{Helstrom76a} that the probability of error between two density
matrices is minimized by performing an optimal positive
operator-valued measure $\mathcal{E}$ (POVM) \cite{Nielsen00a}. The probability of error is defined as \cite{Fuchs99a}
\begin{align}
p_e (\rho_0,\rho_1) \stackrel{{\rm def}}{=} \min_{\mathcal{E} \in \mathcal{M}} p_e (\rho_0 (\mathcal{E}),\rho_1 (\mathcal{E}))
\end{align}
where $\rho_0$ and $\rho_1$ are two arbitrary density matrices and the POVM takes into account all measurements $\mathcal{M}$. Helstrom showed \cite{Helstrom76a} that the probability of error can be expressed explicitly as
\begin{align}
p_e (\rho_0,\rho_1) = \frac{1}{2} + \frac{1}{2} \sum_{\lambda_j \leq 0} \lambda_j
\end{align}
where $\lambda_j$ are the eigenvalues of the matrix $\rho_0 - \rho_1$. It was shown in \cite{Fuchs99a} that the above could be rewritten as
\begin{align}
p_e (\rho_0,\rho_1) = \frac{1}{2} - \frac{1}{4} \sum_{j=1}^N |\lambda_j|
\end{align}
where the summation is over all eigenvalues. Using this, the probability of error can be alternatively expressed as \cite{Helstrom76a}
\begin{eqnarray}
p_e = \frac{1}{2}(1 - D) \label{Error_Probability}
\end{eqnarray}
where $D(\rho_{0},\rho_{1})$ is the trace distance \cite{M.Reed1972,Helstrom76a,Nielsen00a} between the two
density matrices $\rho_{0}$ and $\rho_{1}$ defined as
\begin{eqnarray}\label{distance_measure}
D(\rho_{0},\rho_{1}) &=& \frac{1}{2} {\rm tr} \Big[|\rho_0 - \rho_1|\Big] =
\frac{1}{2}\sum_{j = 1}^{N} |\lambda_{j}|
\end{eqnarray}
Here ${\rm
tr}[|A|]$ is known as the ``trace norm" with $|A| = \sqrt{A^{\dagger} A}$ where $A = \rho_{0} - \rho_{1}$, which has the corresponding eigenvalues $\lambda_{j}$. The distance measure
given here ranges in value from 0, where the two states are
identical, to 1, where the two states are orthogonal, whilst  the
corresponding probability of error ranges from 1/2 to 0, respectively. Also the relation in Eq.~(\ref{Error_Probability}) applies equally to pure or mixed quantum states.
%
For more on the benefits and properties of the trace distance, see e.g., the discussion in \cite{A.Gilchrist2005}. Finally, we point out that in the case of distinguishing between two pure states we have the relation between the trace distance and the fidelity $F$ given by: $D = \sqrt{1 - F}$ \cite{Fuchs99a,A.Gilchrist2005}. In the case of two pure coherent states $\ket{\alpha}$ and $\ket{\beta}$ the probability of error can be written as
\begin{align}\label{eq: helstrom bound}
p_e = \frac{1}{2} (1 - \sqrt{1 - |\bra{\beta}\alpha \rangle|^2})
\end{align}
and is known as the Helstrom bound \cite{Helstrom76a}.

%
%
%
%

%
\begin{figure}[!ht]
\begin{center}
\includegraphics[width=8cm]{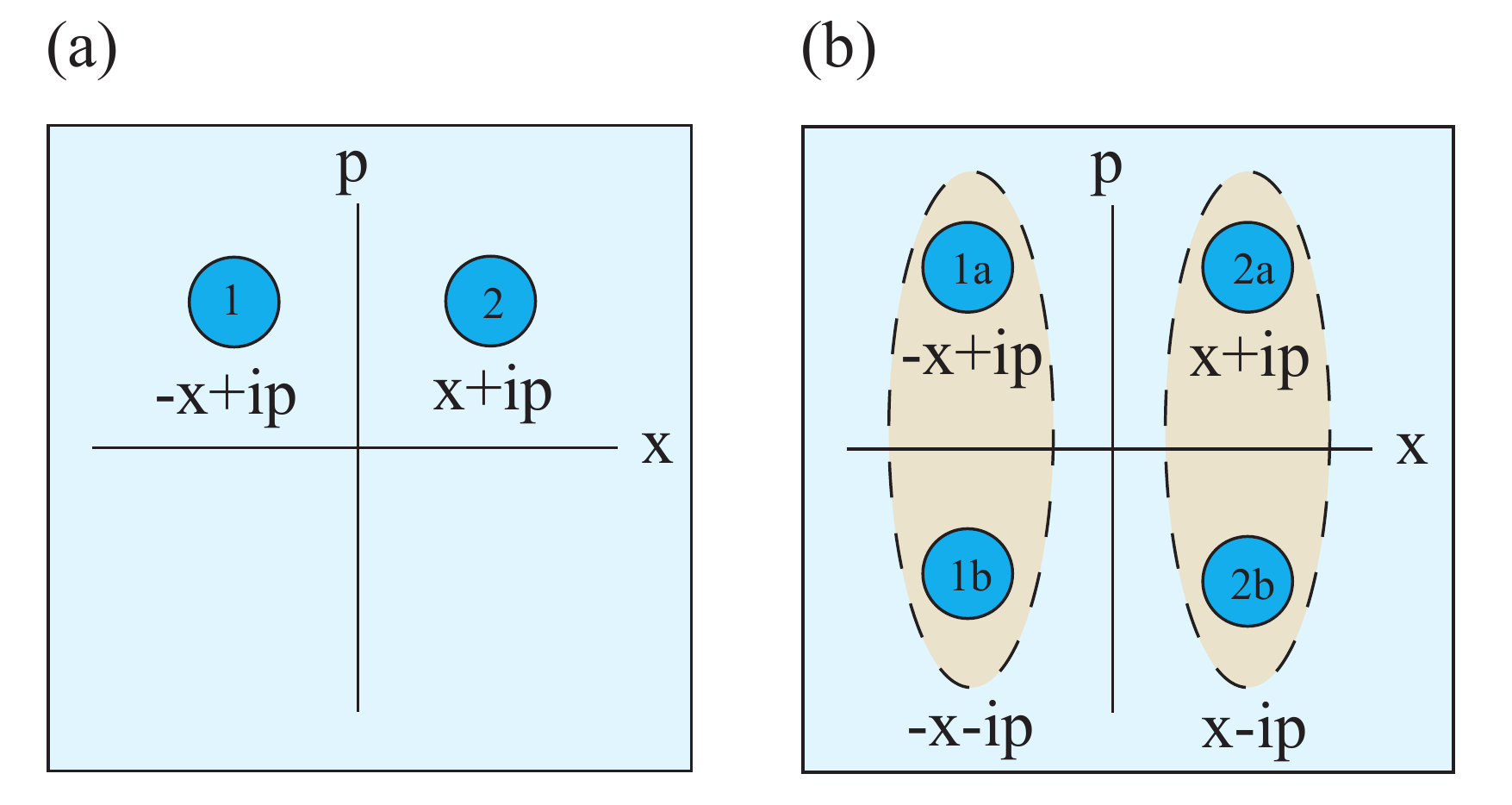}
\caption{Phase-space representation of (a) two pure coherent
states (described by the density operators $\rho_{p_1}$ and $\rho_{p_0}$) and (b) two mixed coherent
states ($\rho_{m_1}$ and $\rho_{m_0}$) for various values of position $x$ and momentum $p$. Here the dotted lines and
shadings in (b) indicate which of the two coherent states are
mixed.}\label{two_mixed_vs_two_pure_discrete}
\end{center}
\end{figure}
\begin{figure}[!ht]
\begin{center}
\includegraphics[width=8cm]{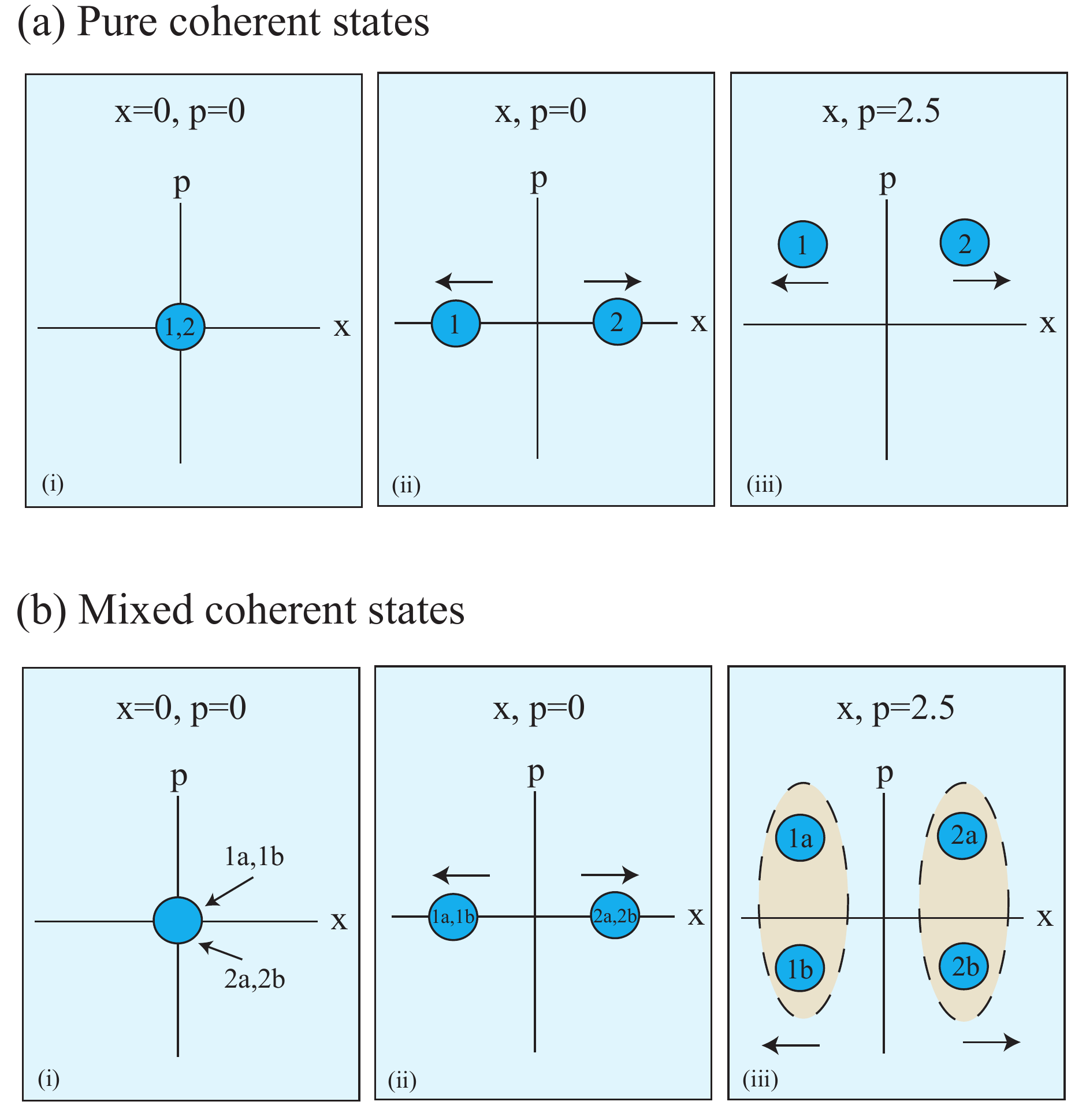}
\caption{Examples of the type of distinguishability situations we consider in this paper. Due to the way that we have set up our phase-space configurations (c.f., Fig.~\ref{two_mixed_vs_two_pure_discrete}) we keep the momentum fixed and vary the position in both the pure state and mixed state cases. For example, in the pure state case (ai) when $x = p = 0$ the two pure states overlap completely and are therefore indistinguishable. (aii) We then keep $p$ fixed and move the pure states further apart by varying $x$. This is then repeated for other fixed values of $p$ (aiii). A similar situation is considered for the mixed state case (b). Here when $p=0$ (bi and bii) we recover the pure state case, whilst for $p \neq 0$, we have the distinguishability of two mixed coherent states (biii).}\label{two_mixed_vs_two_pure_discrete_test_dynamics}
\end{center}
\end{figure}

\section{Distinguishing Pure and Mixed Coherent States}

We will now consider distinguishing between two pure and mixed coherent states using
the previously defined probability of error. A coherent state is
defined as $\ket{\alpha} = D\ket{0}$, where $D = {\rm
exp}(\alpha \hat{a}^{\dagger}-\alpha^* \hat{a})$ is the
displacement operator, and can be written in terms of the Fock state basis as
\begin{align}\label{eq: coherent state}
\ket{\alpha} = {\rm exp}(-\frac{1}{2}|\alpha|^2)
\sum^{\infty}_{n=0} \frac{\alpha^n}{\sqrt{n!}} \ket{n}
\end{align}
A coherent state is also a minimum uncertainty state as well as
an eigenstate of the annihilation operator $\hat{a}$, i.e., $\hat{a} \ket{\alpha} = \alpha \ket{\alpha}$ where $\alpha = x + ip$ is the amplitude of the electromagnetic wave with $\hbar=1/2$.
Any two coherent states $\ket{\alpha}$ and
$\ket{\beta}$ are always non-orthogonal and only approach
orthogonality (i.e., $\bra{\alpha} \beta \rangle \rightarrow 0$)
when $|\alpha-\beta| \gg 1$ where the magnitude is $|\bra{\alpha}
\beta \rangle|^2 = \rm{exp}(-|\alpha-\beta|^2)$. For more background on this, see e.g., \cite{C.C.Gerry2005}. In the following
analysis we will define a coherent state displaced in the amplitude
and phase quadratures, by an amount $x$ and $p$
respectively, as
\begin{align}
\ket{\alpha} \equiv \ket{x + i p}
\end{align}
Consequently, we
can write the density operators of two pure coherent states
$\rho_{p0}$ and $\rho_{p1}$ that we will consider here as
\begin{eqnarray}\label{eq: pure state def}
\rho_{p0} &=& |x + ip\rangle \langle x + ip| \\\nonumber
\rho_{p1} &=& |-x + ip\rangle \langle -x + ip|
\end{eqnarray}
(Note, that in this paper we will interchangeably use the words ``operator" and ``matrix"). In our analysis we also consider two mixed coherent states, where each mixed state is an incoherent mixture of two pure coherent states with equal and opposite displacements in the phase quadrature. The density operators corresponding to these two mixed
states, $\rho_{m0}$ and $\rho_{m1}$, are defined as
\begin{eqnarray}\label{eq: mixed state def}
\rho_{m0} &=& \frac{1}{2} (|x + ip\rangle \langle x + ip| +
|x - ip\rangle \langle x - ip|)
\\\nonumber
\rho_{m1} &=& \frac{1}{2}|-x + ip\rangle \langle -x + ip| +
|-x - ip\rangle \langle -x - ip|)
\end{eqnarray}
Figure~\ref{two_mixed_vs_two_pure_discrete}(a) and
Fig.~\ref{two_mixed_vs_two_pure_discrete}(b) give a
two-dimensional phase-space illustration of the two pure coherent
states and the two mixed coherent states as defined by Eq.~({\ref{eq: pure state def}}) and
Eq.~(\ref{eq: mixed state def}), respectively. Whilst Fig.~\ref{two_mixed_vs_two_pure_discrete_test_dynamics} gives an outline of the type of distinguishability situation we consider for different values of $x$ and $p$.

\begin{figure}[!ht]
\begin{center}
\includegraphics[width=8cm]{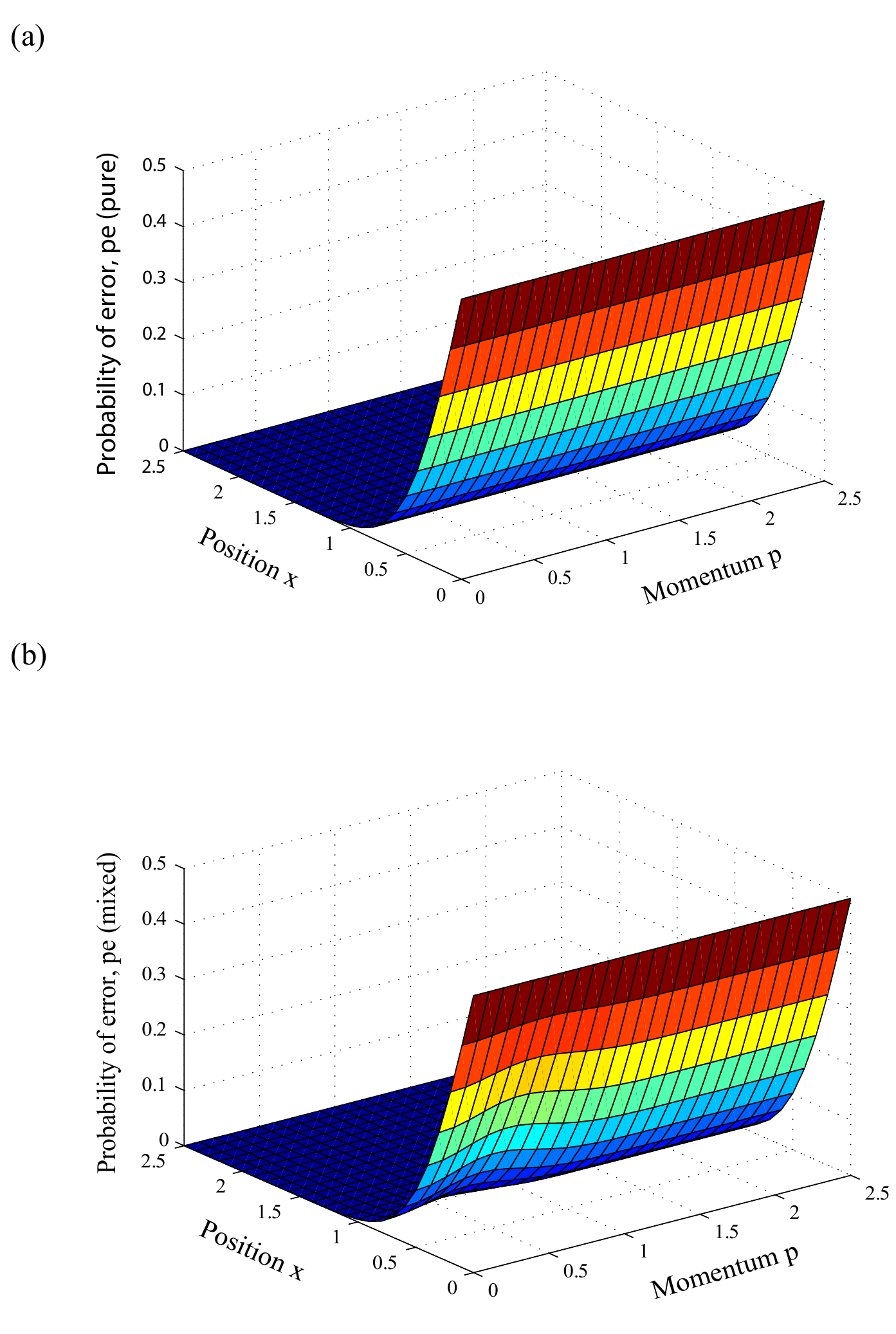}
\caption{Individual plots of the probability of error for (a) two pure
coherent states and (b) two mixed
coherent states using $n=m=50$. Here our values for both position and momentum start at $0$ and go to $2.5$. Both plots exhibit the same overall behavior where the probability of error for the pure state case is independent of the momentum value whilst, for the mixed state case, the probability of error is dependent on certain values of the momentum: $\approx 0 \leq p \leq 1.5$ which alter the probability of error}\label{matlab_plots3}
\end{center}
\end{figure}

According to Eq.~(\ref{distance_measure}) we need to determine
the eigenvalues of $A = \rho_{0} - \rho_{1}$ for
both the two pure states $A^{(p)}$ and the two mixed states $A^{(m)}$, in order to eventually calculate the probability of error. To do this we write
$A$ in its matrix representation which can be expanded in
terms of the Fock state $\ket{n}$ basis defined as
\cite{C.C.Gerry2005}
\begin{equation}
\ket{n} = \frac{(\hat{a}^{\dag})^n}{\sqrt{n!}} \ket{0}
\end{equation}
where $\hat{a}^{\dag}$ is the creation operator of a quantum harmonic
oscillator with $n \in [0,\infty)$. For example, the coherent state
$\ket{x+ip}$ can be written in terms of the Fock state basis using Eq.~(\ref{eq: coherent state}):
\begin{equation}
\ket{x+ip} = e^{-|x+ip|^2/2}\sum_{n=0}^{\infty} \frac{(x+ip)^n}{
\sqrt{n!}}\ket{n}
\end{equation}
Once $A$ is written in matrix form we can then numerically
determine its eigenvalues up to certain values of $n$. First though we want to see what form the matrix elements take. Hence, in this Fock state expansion, the inner
product of an arbitrary coherent state with a Fock state is given by
\begin{eqnarray}
\langle n|\pm x \pm ip \rangle = \frac{(\pm x \pm
ip)^n}{\sqrt{n!}}{\rm exp} (-\frac{1}{2}(x^2 + p^2))\\
\langle \pm x \pm ip|m \rangle = \frac{(\pm x \mp
ip)^m}{\sqrt{m!}}{\rm exp} (-\frac{1}{2}(x^2 + p^2))
\end{eqnarray}
where $\ket{n}$ and $\ket{m}$ are Fock states. Calculating the
general matrix coefficients for the case of two pure coherent
states we
obtain 
\begin{align}\nonumber
\langle n|A^{(p)}|m\rangle &= \frac{{\rm exp}(-x^2 -
p^2)}{\sqrt{n!m!}}[(x+ip)^n(x-ip)^m \\\label{matrices_pure}
&- (-x+ip)^n(-x-ip)^m]
\end{align}
Similarly for the two mixed state case we find
\begin{align}\nonumber
&\langle n|A^{(m)}|m\rangle = \frac{{\rm exp}(-x^2 -
p^2)}{2\sqrt{n!m!}}[(x+ip)^n(x-ip)^m\\\nonumber
&+ (x-ip)^n(x+ip)^m - (-x+ip)^n(-x-ip)^m\\\label{matrices_mixed}
&- (-x-ip)^n(-x+ip)^m]
\end{align}
Numerically we can calculate the eigenvalues of
Eq.~(\ref{matrices_pure}) and Eq.~(\ref{matrices_mixed}) up to
certain values of $n$ and $m$. Then according to
Eq.~(\ref{Error_Probability}) this will give us the probability of
error in distinguishing between two quantum states. These probability of errors are plotted in Fig.~\ref{matlab_plots3} for the pure and mixed state cases using $n=m=50$. We see that both follow the same overall pattern: for various fixed values of momentum $p$, the position $x$ starts from an probability of error of $p_e = 0.5$ when there is no displacement (indistinguishable and hence $50 \%$ chance of guessing the right bit) and tending to $p_e = 0$ after a certain position value: $x \approx 1.5$. We note that a difference between the pure and mixed state cases is the role of $p$. In the pure state case, as expected, the probability of error is the same for any value of $p$ when $x$ is varied. However, there is a small region in the mixed state case $\approx 0 \leq p \leq 1.5$ where for these different values of $p$ the probability of error changes. More specifically, in this region the mixed state probability of error is greater than the pure state probability of error
\begin{align}
p_e^{(m)} > p_e^{(p)}
\end{align}
After $\approx p>1.5$ the two become approximately equal. As we will see, this is what results in the difference in information rates, for certain values of $x$ and $p$, between the pure and mixed states. Now having
numerically calculated $p_e$, we would like to interpret this in
terms of the information gained from using the distinguishing
measure, i.e., the probability of error.

\begin{figure}[!ht]
\begin{center}
\includegraphics[width=8cm]{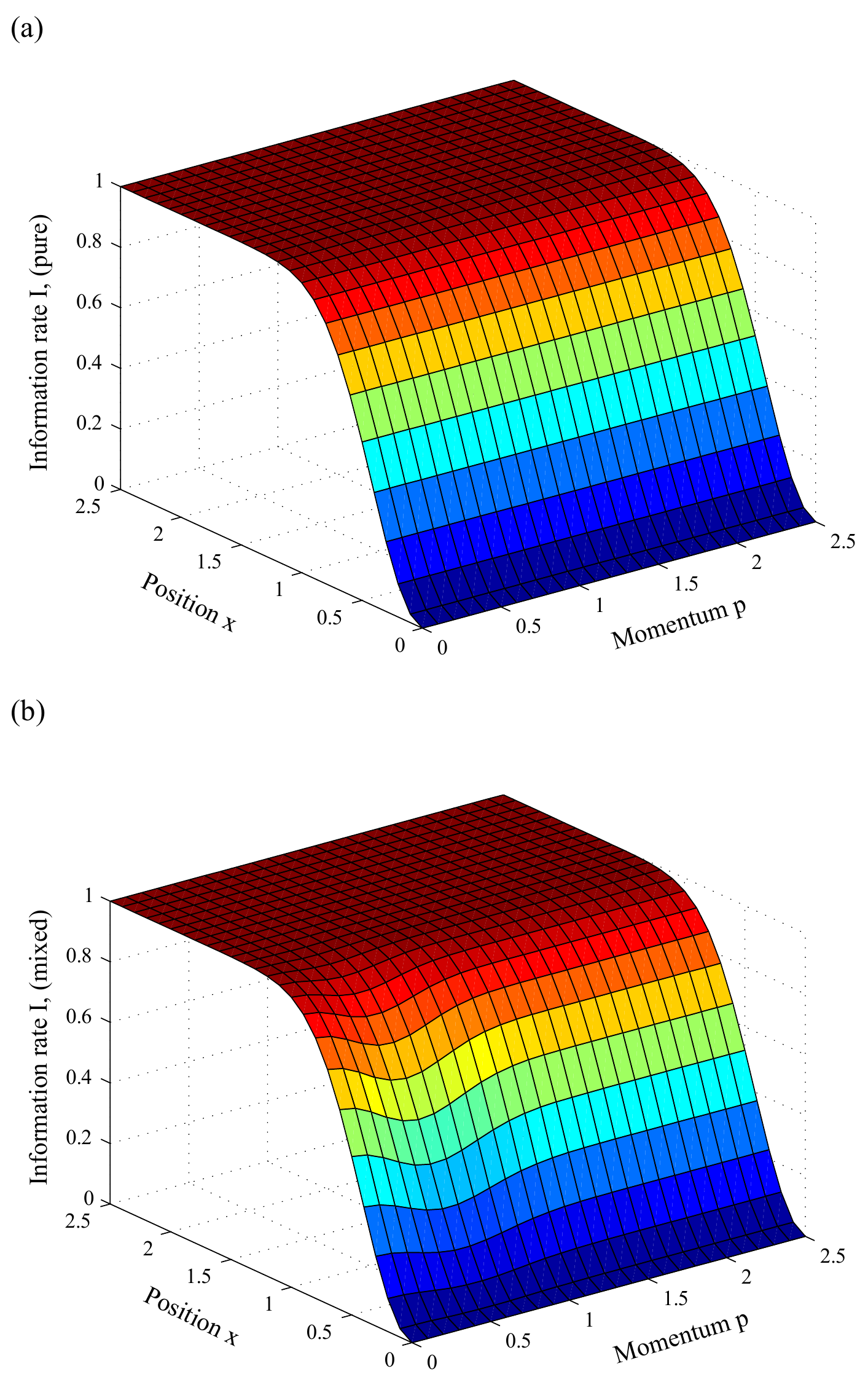}
\caption{Individual plots of the information rate for (a) two pure
coherent states and (b) two mixed
coherent states. Again we have expanded up to $n=m=50$ Fock states in our analysis.}\label{matlab_plots4_new}
\end{center}
\end{figure}
\begin{figure}[!ht]
\begin{center}
\includegraphics[width=8cm]{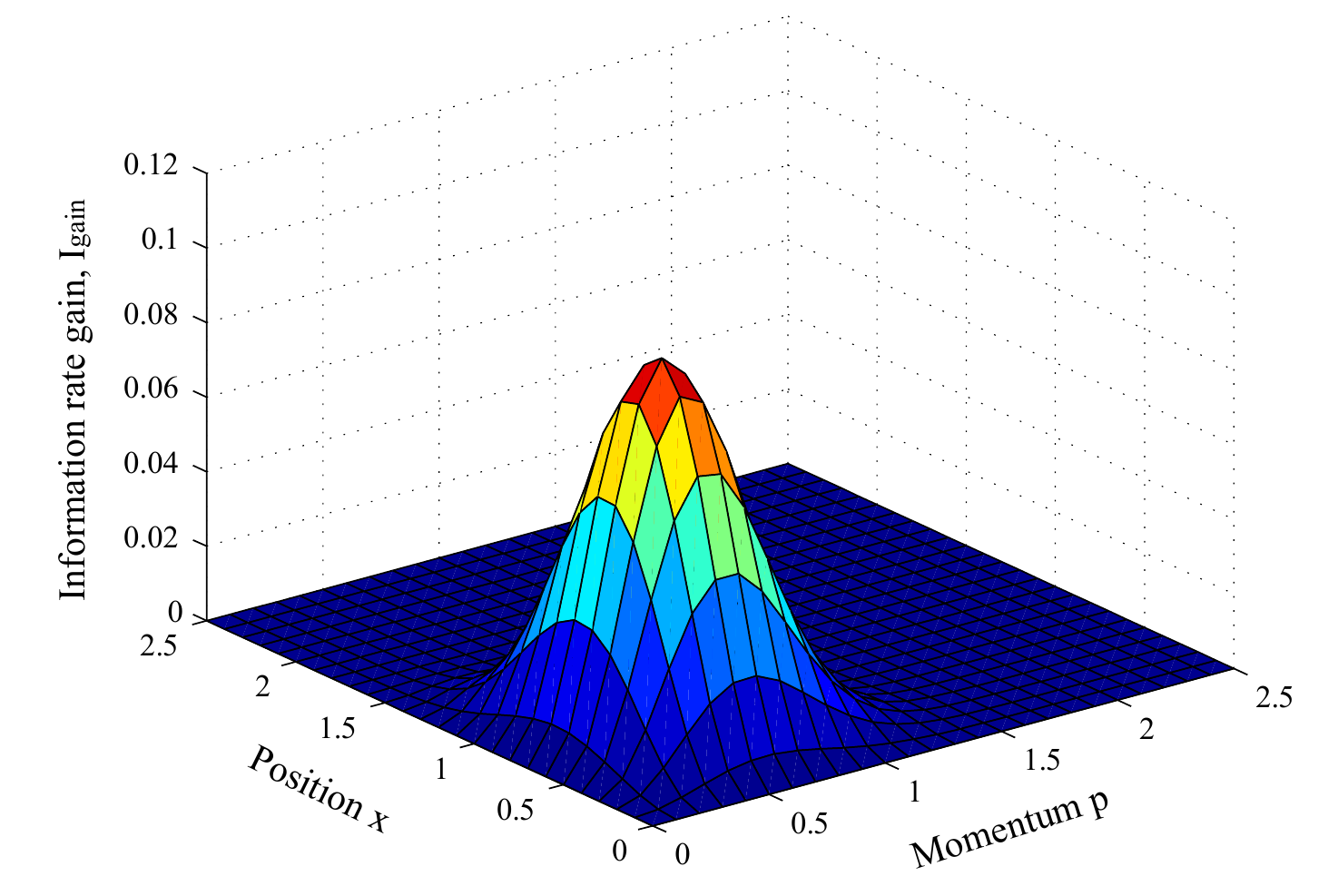}
\caption{The difference in information rates between two pure coherent
states and two mixed coherent states in terms of the position (amplitude) and momentum (phase)
quadratures. Here $I_{gain} = I(\rho_{p_0},\rho_{p_1})-
I(\rho_{m_0},\rho_{m_1})$. For the particular distinguishability case consider here, the two mixed states never
give more information than two pure states which allows us to upper bound information rates in CV-QKD using the post-selection protocol.}\label{3D_plot}
\end{center}
\end{figure}

\subsection{Shannon Information}

The information obtained by distinguishing between two states can be calculated
using the well known Shannon information formula for a binary symmetric
channel \cite{Shannon48a}
\begin{eqnarray}\label{shannon_formula}
I = 1 + p_e {\rm log_{2}}p_e + (1 - p_e){\rm log_{2}}(1 - p_e).
\end{eqnarray}
Figure~\ref{3D_plot} shows the difference between the Shannon
information obtained by distinguishing between two coherent states
$I(\rho_{p_0},\rho_{p_1})$ compared with distinguishing between
two mixed coherent states $I(\rho_{m_0},\rho_{m_1})$ (where the individual cases are plotted in Fig.~\ref{matlab_plots4_new}). This information
difference is defined as the information gain $I_{gain}$
\begin{align}
I_{gain} = I(\rho_{p_0},\rho_{p_1})-
I(\rho_{m_0},\rho_{m_1})
\end{align}
Figure~\ref{3D_plot} plots $I_{gain}$ in terms of the position (amplitude) and momentum (phase) quadrature displacements of
the pure and mixed states as defined in Eq.~(\ref{eq: pure state def}) and Eq.~(\ref{eq: mixed state def}),
respectively. Here we have expanded up to $50$ Fock states, i.e.,
$n = m = 50$ in our numerical analysis.

There are two main features of Fig.~\ref{3D_plot}. Firstly, we notice that,
given our distinguishability measure and initial configuration of
coherent states in phase-space, two mixed states never give more
information than two pure states, i.e.,
\begin{equation}
I(\rho_{m0},\rho_{m1}) \leq
I(\rho_{p0},\rho_{p1})
\end{equation}
This result is relevant given that in the original post-selection CV-QKD protocol \cite{Silberhorn2002} it means that an eavesdropper is upper bounded, in terms of her accessible information, when choosing to distinguish between two pure coherent states (instead of the two mixed coherent states). Secondly, there is a flat region where the information gain is
zero, i.e., $I_{gain} = 0$ where the information from distinguishing between two mixed
states is the same as that of two pure states. This means that as the pure coherent states and the mixed coherent states are moved further and further apart in the amplitude
quadrature (for fixed values of momentum), the
probability of error tends to the same value (i.e., $p_e = 0$) and hence the same amount of information is obtained from both (c.f., Fig.~\ref{matlab_plots3}). So in some sense,
at a certain point the two mixed states start ``behaving" (from a distinguishability point of view) like two pure states. This only starts occurring for the mixed states when the value of $p$ is greater than a particular value. This is because we require that the individual mixed states themselves are further apart (in $p$ value) and hence more distinguishable individually, before we can then start distinguishing each of the two mixed states with one other.

\subsubsection{Discussion: Maximum Accessible Information for an Eavesdropper}

We now briefly note the equivalence between the information rate obtained using the probability of error as defined in~Eq.~(\ref{Error_Probability}) and the Levitin information bound \cite{Levitin95a} which is used in post-selection CV-QKD to ascertain how much information an eavesdropper gains. The original post-selection protocol \cite{Silberhorn2002} involves the generation of a secure key by Alice sending Bob coherent states that have had classical variables $\{x,p\}$ encoded on them. Bob measures these states using homodyne (or heterodyne \cite{Lance2005}) detection and then decodes them using some previously agreed upon binary encoding. To calculate how much information an eavesdropper can optimally obtain during the protocol we use the Levitin bound \cite{Levitin95a} which determines the maximum accessible information from distinguishing between two non-orthogonal pure states, i.e.,
\begin{eqnarray}\nonumber
I_{AE} &= \frac{1}{2}(1 + \sqrt{1 - |z|^2}){\rm log}_2 (1 + \sqrt{1 -
|z|^2})\\\label{eq: levitin bound original}
&+ \frac{1}{2}(1 - \sqrt{1 - |z|^2}){\rm log}_2 (1 -\sqrt{1 -
|z|^2})
\end{eqnarray}
where $I_{AE}$ denotes the mutual information between Alice and the eavesdropper, Eve. Here $z$ is the overlap of the two pure coherent states which Eve needs to distinguish between, i.e.,  $z = \langle~-~x~+~i~p~\ket{x+ip} = {\rm exp} [-2 (x^2 + ixp)]$ \cite{C.C.Gerry2005} where the modulus squared is the Gaussian $|z|^2 = {\rm exp} (-4x^2)$.
Again we assumed that the channel transmission is set to unity. We note here that Eq.~(\ref{eq: levitin bound original}) can be alternately derived by simply using the probability of error given by the Helstrom bound for two pure coherent states, i.e.,  Eq.~(\ref{eq: helstrom bound}), and substituting that into Shannon's formula given in Eq.~(\ref{shannon_formula}). Consequently, after some simple algebra, we see that $I(\rho_{p0}, \rho_{p1}) = I_{AE}$. This result only applies to the pure state case. The question of maximizing the accessible information in CV quantum state discrimination (and hence, in CV-QKD eavesdropping analysis) for two general mixed quantum states is still an open question. Although Levitin does discuss a specific (non-general) situation in \cite{Levitin95a}.



\subsection{Homodyne Detection versus POVM: Pure and Mixed State Cases}

In this section we consider the following questions: what is the probability of error in distinguishing between two pure coherent states and two mixed coherent states (whose orientation is defined in Fig.~\ref{two_mixed_vs_two_pure_discrete}) given that a homodyne detection (also known as a projective or von Neumann) measurement is performed? And how does that compare to the probability of error defined using the trace distance? Homodyne detection is one of the most commonly used methods of measurement in CV quantum communication protocols~\cite{S.L.Braunstein2005}, and consequently, these questions are of practical interest, particularly for CV-QKD. We note that previous work on this includes binary optical communication distinguishability using direct and homodyne detection in realistic situations \cite{Oli04}. As well as optical pure coherent state distinguishability which has been theoretically \cite{Dol73,Ken73,Tak05,Sas96,Tak08,Ger04} and experimentally investigated \cite{Coo07,Wit08}. However, in the following analysis we consider a specific distinguishability situation for both the pure and mixed coherent state cases as given in Fig.~\ref{two_mixed_vs_two_pure_discrete} which is motivated by post-selection CV-QKD.

\begin{figure}[!ht]
\begin{center}
\includegraphics[width=8cm]{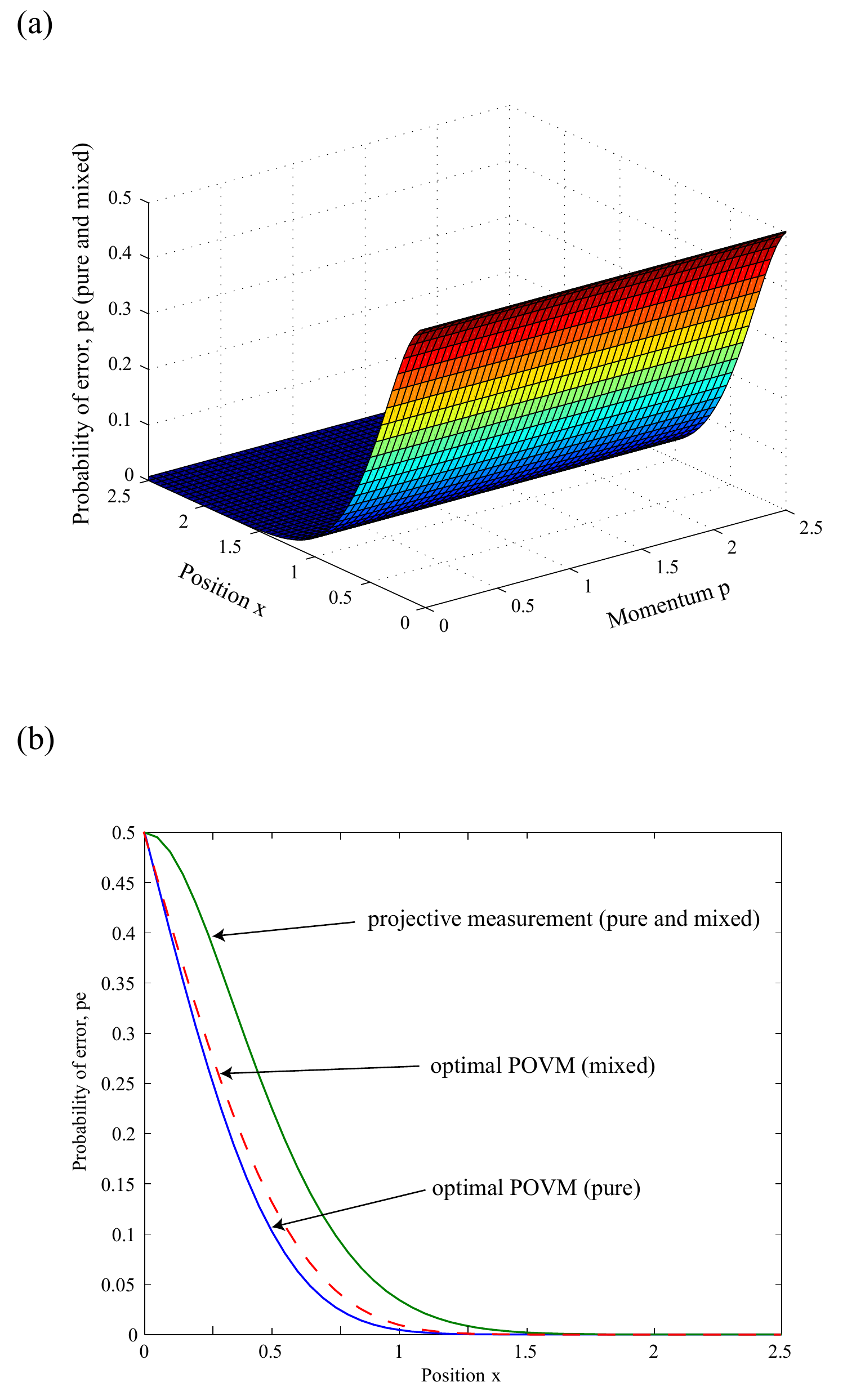}
\caption{Probability of error plots. (a) The plot of the probability of error for both the pure and mixed coherent states for an $x$ quadrature measurement. Both are identical due to the fact that now a projective (homodyne) measurement is performed on each of the quantum states rather than the usual POVM as considered before. (b) Optimal POVMs versus projective measurements for the pure and mixed coherent state cases. In Figs.~\ref{matlab_plots3}~(a) and \ref{figurePOVM}~(a), due to probability of error being independent of the momentum variable $p$, we can take a cross-sectional slice at any value of $p$ and plot the probability of error as a function of $x$ for both types of measurements. We also plot a cross-sectional slice from the mixed state case given in Fig.~\ref{matlab_plots3}(b) for $p=0.55$ (red dashed line). As expected the POVM minimizes the probability of error, i.e., the projective measurement is not the optimal type of distinguishability measurement.}\label{figurePOVM}
\end{center}
\end{figure}

We will first analyze the pure state case as the results for both the pure state and mixed state cases will be the same. The reason for this can be seen from Fig.~\ref{two_mixed_vs_two_pure_discrete} where an $\hat{x}$ quadrature measurement will collapse and project the mixed states onto the $x$ axis in the same way as the pure state case does. An $\hat{x}$ quadrature measurement using homodyne detection is modeled theoretically by acting a projective measurement $\ket{x}\bra{x}$ on the two pure coherent states $\ket{x+ ip}$ and $\ket{-x+ ip}$. The probability of obtaining the measurement outcome $m$ is given by \cite{U.Leonhardt1997,Oli04}
\begin{align}
P(m|\ket{\pm x+ip}) &= |\langle x|\pm x+ip \rangle|^{2} = \sqrt
\frac{2}{\pi} e^{-2(m \mp x)^{2}}
\label{probability_distribution_original}
\end{align}
Such a formula is used to derive information rates for (the receiver) Bob in the CV-QKD post-selection protocol, except in the above formula the loss on the quantum channel $\eta$ (which is typically associated with the eavesdropper) is set to unity $\eta = 1$. The probability of error $p_e^{(p)}$, when a projective measurement is performed to distinguish between the two pure coherent states, can now be written as
\begin{eqnarray}
p_{e}^{(p)} = \left\{ \begin{array}{ll} \frac{P(m|\ket{-x + ip})}{P(m|\ket{x + ip})
+ P(m|\ket{-x + ip})} & \textrm{for $m > 0$}\\\\
\frac{P(m|\ket{x + ip})}{P(m|\ket{x + ip}) + P(m|\ket{-x + ip})} & \textrm{for $m <
0$}
\end{array} \right.\label{pe_original}
\end{eqnarray}
Substituting Eq.~(\ref{probability_distribution_original}) into Eq.~(\ref{pe_original}) leads to
\begin{align}
p_{e}^{(p)} = \left\{ \begin{array}{ll} \frac{e^{-2(m + x)^{2}}}{e^{-2(m - x)^{2}}+ e^{-2(m + x)^{2}}} & \textrm{for $m > 0$}\\\\
\frac{e^{-2(m - x)^{2}}}{e^{-2(m - x)^{2}}+ e^{-2(m + x)^{2}}} & \textrm{for $m <0$}
\end{array} \right.\label{pe_original_sub_values}
\end{align}
%
The final probability of error $\bar{p}_e^{(p)}$ once we have integrated over all possible measurement results $m$ is given by
\begin{align}\nonumber
\bar{p}_e^{(p)} &= 2 \int_0^{\infty} dm \hspace{1mm} p_e^{(p)} P(m|\ket{x+ip})\\
&= \sqrt{\frac{8}{\pi}} \int_0^{\infty} \frac{dm}{e^{2(m + x)^2} +  e^{2(-m + x)^2}}
\end{align}
We numerically evaluate the above integral and plot the results in Fig.~\ref{figurePOVM}~(a). Again the resulting plots for both the pure state case and the mixed state case are identical. Fig.~\ref{figurePOVM}~(a) has similar behavior as that of the probability of error obtained using the trace distance for the pure coherent state case, i.e., Fig.~\ref{matlab_plots3}~(a). Because the probability of errors given in Figs.~\ref{matlab_plots3}~(a) and \ref{figurePOVM}~(a)) are independent of the value of momentum we can plot a 2-D cross-sectional slice of the probability of error for both the POVM and projective measurement cases. This is given in Fig.~\ref{figurePOVM}~(b) where it can be seen that the measurement associated with the probability of error using a POVM (i.e., as a function of the trace distance) is lower than the projective $\hat{x}$ quadrature homodyne detection measurement. The distance between the two outside curves in Fig.~\ref{figurePOVM}~(b) is slightly reduced for certain values of $p$ when considering the mixed state case. For example, in Fig.~\ref{matlab_plots3}~(b) for values where $0 < p \leq 1.5$ the probability of error is slightly increased. We illustrate this in Fig.~\ref{figurePOVM}~(b) by plotting the mixed state case for $p=0.55$.

\section{Distinguishing Pure and Mixed Squeezed States}

\begin{figure}[!ht]
\begin{center}
\includegraphics[width=8cm]{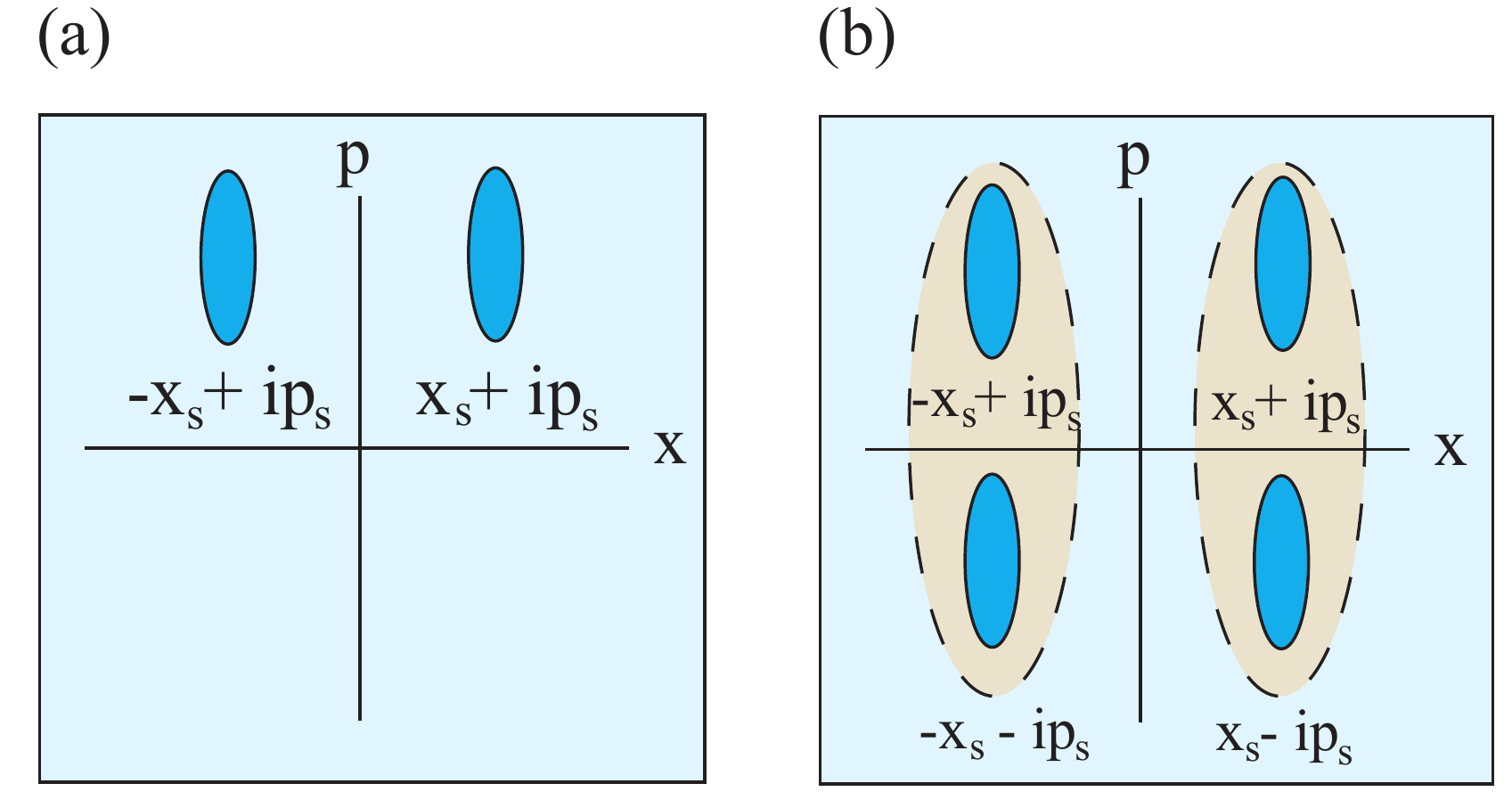}
\caption{Phase-space representation of (a) two pure squeezed
states (described by the density operators $\rho_{p_1}$ and $\rho_{p_0}$) and (b) two mixed squeezed
states ($\rho_{m_1}$ and $\rho_{m_0}$) for arbitrary displacements $\{x_s,p_s\}$. Again, just as with the coherent state configuration, the dotted lines and
shadings in (b) indicate which of the two squeezed states form a
mixture.}\label{two_mixed_vs_two_pure_discrete_test_SQZ}
\end{center}
\end{figure}

Having analyzed the distinguishability of pure and mixed coherent states, we now extend our analysis to another set of Gaussian states: squeezed states. Figure~\ref{two_mixed_vs_two_pure_discrete_test_SQZ} gives a phase-space representation of the distinguishability situation we consider, i.e., it is the same configuration as the coherent state case except now we are considering it for displaced squeezed states. A displaced squeezed vacuum state \cite{C.C.Gerry2005} is defined as
\begin{align}
\ket{\alpha, \xi} = D S \ket{0}
\end{align}
where $S$ is the single-mode squeezed gate defined as
\begin{align}
S = {\rm exp} [r(\hat{a}^2 - \hat{a}^{\dagger 2})/2] = {\rm exp} [ir(\hat{x}\hat{p} + \hat{p}\hat{x})]
\end{align}
and $r$ is the squeezing parameter ($r \in [0,\infty$)) performed in only one (position) quadrature direction and again $D$ is the displacement gate. As can be seen, the above state is created by first squeezing the vacuum state $\ket{0}$ and then displacing it. We will now define the density operators of two pure squeezed states that we consider as
\begin{eqnarray}
\rho_{p0} &=& |x_s + ip_s\rangle \langle x_s + ip_s| \\
\rho_{p1} &=& |-x_s + ip_s\rangle \langle -x_s + ip_s|
\end{eqnarray}
where the subscript $s$ indicates that we are now considering the squeezed state situation. We can define the density operators of two arbitrary mixed
squeezed states $\rho_{m0}$ and $\rho_{m1}$ as
\begin{align}
\rho_{m0} &= \frac{1}{2} (|x_s + ip_s\rangle \langle x_s + ip_s| +
|x_s - ip_s\rangle \langle x_s - ip_s|)
\\\nonumber
\rho_{m1} &= \frac{1}{2} \Big(|-x_s + ip_s\rangle \langle -x_s +
ip_s|\\
&+ |-x_s - ip_s\rangle \langle -x_s - ip_s| \Big)
\end{align}
\begin{figure}[!ht]
\begin{center}
\includegraphics[width=8cm]{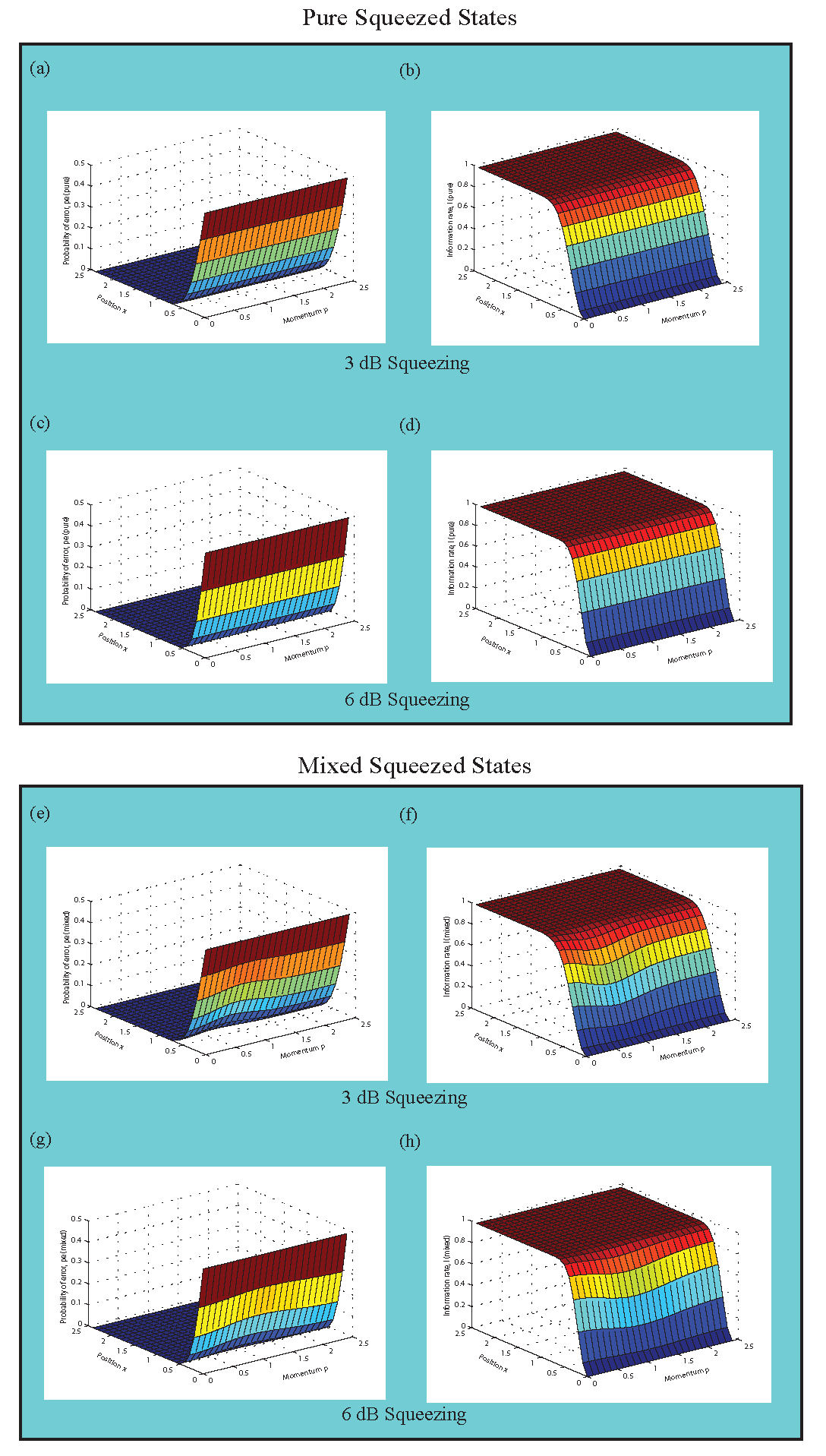}
\caption{Individual plots of both the probability of errors and information rates for two pure squeezed states (a) to (d) and two mixed squeezed states (e) to (h) for two types of squeezing parameters: $r = 0.35$ ($3$ dB) and $r = 0.70$ ($6$ dB). These plots reflect the same overall behavior and characteristics of the probability of error and information rates which were exhibited in the coherent state case. The chief difference is the increase in squeezing results in the distinguishability measure $p_e$ tending towards zero for smaller values of position $x$.}\label{matlab_plots_SQZ1}
\end{center}
\end{figure}
\begin{figure}[!ht]
\begin{center}
\includegraphics[width=8cm]{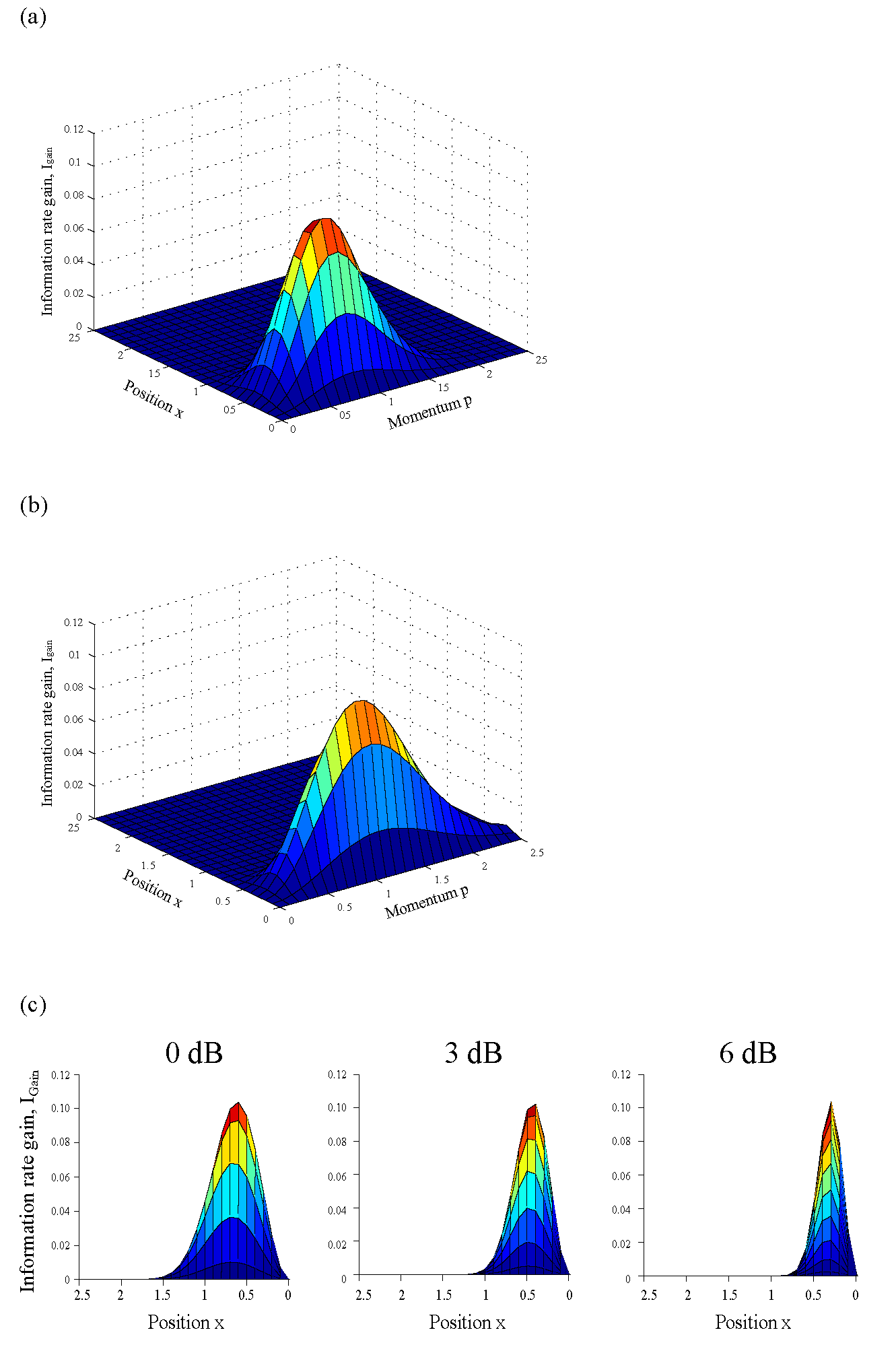}
\caption{The difference in information rates $I_{gain}$ between the two pure squeezed
states and two mixed squeezed states for two types of squeezing: (a) $r = 0.35$ ($3$ dB) and (b) $r = 0.70$ ($6$ dB). As with the coherent state case, two mixed squeezed states never give more information than two pure squeezed states, with respect to the phase-space configurations considered in this paper. (c) Side-on profile of the variation in the information distribution for the three cases studied: $0$ dB (coherent state, c.f., Fig.~\ref{3D_plot}), $3$ dB and $6$ dB (squeezed states).}\label{3D_plot_SQZ3}
\end{center}
\end{figure}
As with the coherent state analysis, we need to ultimately determine the eigenvalues of the appropriate matrices in order to calculate the trace distance and then the probability of error. Again this involves expanding the matrix in an orthogonal basis, i.e., the Fock state basis. With this in mind, a displaced squeezed state can be expanded in terms of Fock states as \cite{C.C.Gerry2005}
\begin{align}\label{eq: sqz fock state}
\ket{\alpha,\xi} &= \frac{1}{\sqrt{{\rm cosh} r}} {\rm exp}
\Big[-\frac{1}{2}(|\alpha|^2 + \alpha^{*2} e^{i \theta} {\rm tanh
}r) \Big]\\\nonumber
&\times \sum^{\infty}_{n=0}
\frac{\Big[\frac{1}{2} e^{i \theta} {\rm tanh}
r \Big]^{n/2}}{\sqrt{n!}} H_n \Big[\gamma (e^{i \theta}{\rm
sinh}2r)^{-1/2} \Big] \ket{n}
\end{align}
where $\alpha = x_s + ip_s$, $\xi = r e^{i \theta}$ and $\gamma = \alpha {\rm cosh} r + \alpha^* e^{i \theta} {\rm sinh}
r$. Here $H_n(x)$ are the Hermite
polynomials of degree $n$ which are a polynomial sequence defined as
\begin{eqnarray}
H_n(x) = (-1)^n e^{x^2} \frac{d^n}{dx^n} e^{-x^2}
\end{eqnarray}
where in our case $x \equiv \gamma (e^{i \theta}{\rm sinh}2r)^{-1/2}$. We point out that in the limit $r \rightarrow 0$ in Eq.~(\ref{eq: sqz fock state}) we simply get back the coherent state as given by Eq.~(\ref{eq: coherent state}). Note that in our calculations we will only consider the case when $\theta=0$, i.e., the squeezed states are only squeezed along the position quadrature, c.f., Fig.~\ref{two_mixed_vs_two_pure_discrete_test_SQZ}, and not along some angle $\theta$.

Using Eq.~(\ref{eq: sqz fock state}) the overlap of a Fock state and a displaced squeezed state is given by
\begin{align}\nonumber
\langle n\ket{\alpha,\xi} &= (n! {\rm cosh} r)^{-1/2} {\rm exp}
\Big[-\frac{1}{2}(|\alpha|^2 + \alpha^{*2} e^{i \theta} {\rm tanh}
r) \Big]\\
&\times
\Big[\frac{1}{2} e^{i \theta} {\rm
tanh} r \Big]^{n/2} H_n \Big[\gamma (e^{i \theta}{\rm
sinh}2r)^{-1/2} \Big]
\end{align}
We also have for the other matrix elements
\begin{align}\nonumber
\langle \alpha,\xi\ket{m} &= (m! {\rm cosh} r)^{-1/2} {\rm exp}
\Big[-\frac{1}{2}(|\alpha|^2 + \alpha^{2} e^{-i \theta} {\rm tanh}
r) \Big]\\
&\times \Big[\frac{1}{2} e^{-i \theta} {\rm
tanh} r \Big]^{m/2} H_m \Big[\gamma^* (e^{-i \theta}{\rm
sinh}2r)^{-1/2} \Big]
\end{align}
Writing the above in the $x$ and $p$ notation gives us
\begin{align}\nonumber
&\langle n\ket{x_s+ip_s} = (n! {\rm cosh} r)^{-1/2} {\rm exp}
\Big[-\frac{1}{2}(x_s^2+p_s^2 + (x_s-ip_s)^2\\
&\times e^{i \theta} {\rm tanh}
r) \Big] \Big[\frac{1}{2} e^{i \theta} {\rm
tanh} r \Big]^{n/2} H_n \Big[\gamma (e^{i \theta}{\rm
sinh}2r)^{-1/2} \Big]
\end{align}
where $\gamma = (x_s+ip_s) {\rm cosh} r + (x_s-ip_s) e^{i \theta} {\rm sinh}
r$. We also have
\begin{align}\nonumber
&\langle x_s+ip_s\ket{m} = (m! {\rm cosh} r)^{-1/2} {\rm exp}
\Big[-\frac{1}{2}(x_s^2+p_s^2 + (x_s+ip_s)^2\\
&\times
e^{-i \theta} {\rm tanh}
r) \Big] \Big[\frac{1}{2} e^{-i \theta} {\rm
tanh} r \Big]^{m/2} H_m \Big[\gamma^* (e^{-i \theta}{\rm
sinh}2r )^{-1/2} \Big]
\end{align}
with $\gamma^* = (x_s-ip_s) {\rm cosh} r + (x_s+ip_s) e^{-i \theta} {\rm sinh}
r$.
Again calculating the general matrix coefficients for the case of the two pure squeezed
states we obtain
\begin{align}\nonumber
&\langle n|A_s^{(p)}|m\rangle = \frac{(n!m!)^{-1/2}}{{\rm cosh} r} \Big(\frac{1}{2} {\rm tanh} r \Big)^{\frac{n+m}{2}} {\rm exp}[-(x_s^2 +p_s^2\\\label{matrices_pure_sqz}
& + {\rm tanh} r (x_s^2-p_s^2))] [H_n (\gamma) H_m (\gamma^*) - H_n (\gamma') H_m (\gamma^{*'})]
\end{align}
where $A_s^{(p)} = \rho_{p0} - \rho_{p1}$ and $\theta = 0$. Here the Hermite polynomials are defined as:
\begin{align}\nonumber
H_n(\gamma) &\equiv H_n[\gamma ({\rm
sinh}2r)^{-1/2}]\\\nonumber
H_m(\gamma^*) &\equiv H_m[\gamma^* ({\rm
sinh}2r)^{-1/2}]\\\nonumber
H_n(\gamma') &\equiv H_n[\gamma' ({\rm
sinh}2r)^{-1/2}]\\\label{eq: hermite poly}
H_m(\gamma^{'*}) &\equiv H_m[\gamma^{'*} ({\rm
sinh}2r)^{-1/2}]
\end{align}
where $\gamma$ is defined as usual (but now with $\theta =0$) and $\gamma^{'} = (-x_s + ip_s) {\rm cosh} r + (-x_s - ip_s) {\rm sinh} r$. Similarly for the two mixed squeezed states we find the matrix elements are given by
\begin{align}\nonumber
&\langle n|A_s^{(m)}|m\rangle = \frac{1}{2} \frac{(n!m!)^{-1/2}}{{\rm cosh} r} \Big(\frac{1}{2} {\rm tanh} r\Big)^{\frac{n+m}{2}} \times \\\nonumber
&\times {\rm exp}[-(x_s^2 +p_s^2 + (x_s^2-p_s^2) {\rm tanh} r)] [H_n (\gamma) H_m (\gamma^*) +\\\label{matrices_mixed_sqz}
&+ H_n (\gamma'') H_m (\gamma^{''*})- H_n (\gamma') H_m (\gamma^{'*}) - H_n (\gamma''') H_m (\gamma^{'''*})]
\end{align}
where $A_s^{(m)} = \rho_{m0} - \rho_{m1}$ and the Hermite polynomials are defined previously in Eq.~(\ref{eq: hermite poly}) with the new additional variables
\begin{align}
\gamma^{''} &= (x_s-ip_s) {\rm cosh} r + (x_s+ip_s) {\rm sinh} r\\
\gamma^{'''} &= (-x_s-ip_s) {\rm cosh} r + (-x_s+ip_s) {\rm sinh} r
\end{align}
plus their respective conjugates. We can simplify the Hermite polynomials in Eq.~(\ref{matrices_mixed_sqz}) by realizing that the following relations hold
\begin{align}
\gamma'' = \gamma^* \hspace{1cm} \gamma''' = \gamma^{'*}
\end{align}
plus their conjugates. Therefore Eq.~(\ref{matrices_mixed_sqz}) can be rewritten as:
\begin{align}\nonumber
&\langle n|A_s^{(m)}|m\rangle = \frac{1}{2} \frac{(n!m!)^{-1/2}}{{\rm cosh} r} \Big(\frac{1}{2} {\rm tanh} r \Big)^{\frac{n+m}{2}} \times\\\nonumber
&\times {\rm exp}[-(x_s^2 +p_s^2 +  (x_s^2-p_s^2) {\rm tanh} r)] \times [H_n (\gamma) H_m (\gamma^*) +\\\label{matrices_mixed_sqz_simpl}
&+ H_n (\gamma^*) H_m (\gamma)- H_n (\gamma') H_m (\gamma^{'*}) - H_n (\gamma^{'*}) H_m (\gamma^{'})]
\end{align}
Numerically we can calculate the eigenvalues of
Eq.~(\ref{matrices_pure_sqz}) and Eq.~(\ref{matrices_mixed_sqz_simpl}) for two values of the squeezing parameter, $r$. According to Eq.~(\ref{Error_Probability}) this will give us the probability of error in distinguishing between the two sets of quantum states. Our results are plotted in Fig.~\ref{matlab_plots_SQZ1}(a), (c), (e), and (g) where we have set $\theta=0$ and used two squeezing parameters: (1) $r=0.35$ which corresponds to approximately 3 dB of squeezing and (2) $r = 0.7$ ($6$ dB) (these conversions are obtained by using the formula: $10 {\rm log}_{10} (e^{-2r})$ dB). We can see that as the squeezing is increased the probability of error is reduced in both the pure and mixed state cases. For example, in the coherent state case for fixed values of momentum $p_e \rightarrow 0$ when $\approx x > 1.5$. However in the pure squeezed state case, the position value is $\approx x > 1$ for $3$ dB of squeezing and $\approx x > 0.75$ for $6$ dB. The reason this occurs can be seen by comparing the distinguishability of two pure coherent states with two pure squeezed states. If you first picture the two coherent states initially overlapping (at $x=p=0$, c.f., Fig.~\ref{two_mixed_vs_two_pure_discrete_test_dynamics}~(a)) and then increasing the $x$ distance between them to a point where the phase-space circles no longer overlap. Now doing this again but with the $x$ quadrature squeezed states, we can see that because these circles are narrower then it takes a smaller distance for them to no longer overlap. Hence, a smaller $x$ is required to achieve a smaller probability of error.

\subsection{Shannon Information}

Again we calculate the Shannon information to obtain the information rate gain $I_{gain}$ for the two values of squeezing and plot them in Fig.~\ref{3D_plot_SQZ3} (where the individual rates are given in Fig.~\ref{matlab_plots_SQZ1}~(b), (d), (f), and (h)). As with the coherent state analysis, based on our distinguishability measure and initial configuration in phase-space, two mixed squeezed states (where each mixed state is an incoherent mixture of two pure squeezed states with equal and opposite displacements in the phase quadrature) never give more information than two pure squeezed states. We again see, after certain values of position and momentum, a flat region in both graphs which indicates that the two mixed states have the same accessible information as the two pure states. This results in a net information rate of $I_{gain} = 0$ and is due to the same reason as was given for the coherent states. The effect of increasing the squeezing parameter is given in Fig.~\ref{3D_plot_SQZ3}(a) and (b) with a side-on profile depicted in Fig.~\ref{3D_plot_SQZ3}(c). Figure~\ref{3D_plot_SQZ3}(c) shows that by increasing the amount of squeezing in the $x$ direction the effect for fixed $p$ is to narrow the information distribution for different position values. In some sense we have a ``squeezing" of the information rate along the $x$ axis. This comes from the fact that, as mentioned before, as the squeezing increases the probability of error decreases for smaller values of $x$. This ultimately leads to $I_{gain} \rightarrow 0$ for smaller values of $x$ than what we had for the (zero squeezing) coherent state case (c.f., Fig.~\ref{3D_plot_SQZ3}(c)). Conversely, we also notice that as squeezing is increased, for fixed values of $x$, the net information rate requires larger values of $p$ until $I_{gain} = 0$. This leads to a broadening or an ``anti-squeezing" of the information rate along the $p$ direction.

\section{Discussion and Conclusion}

In our analysis we used the probability of error to discriminate between specific phase-space configurations of two pure and two mixed CV quantum states. Recently \cite{S.Pirandola2008b}, Pirandola and Lloyd combined the Minkowski inequality and the quantum Chernoff bound to derive upper bounds on quantum state discrimination for CV. This was in the context of Gaussian states using symplectic algebraic methods (e.g., see \cite{S.Pirandola2008}). Future work would entail comparing these techniques to the ones given in this paper and extending it to include other Gaussian states, such as EPR states and thermal states.

In conclusion, we have considered a situation in post-selection based CV-QKD where there is an assumption that an eavesdropper upper bounds her information by distinguishing between two pure coherent states instead of distinguishing between two
mixed coherent states (where the various mixtures have the same
position component). We showed that the eavesdropper will never get more
information from the two mixed coherent states. Hence, we have proven the assumption to be true. We showed this
using the probability of error as the distinguishability measure
along with the Shannon information formula. Furthermore, we expanded our analysis to include other types of Gaussian states: pure and mixed squeezed states. In that analysis, the squeezed states are aligned in phase-space in the same configuration as the coherent states were. The same types of behavior and characteristics are present in the probability of error and information rate plots for the squeezed states as was for the coherent state case. Furthermore, varying the amount of squeezing results in the ``squeezing" and ``anti-squeezing" of the net information gain rates, i.e., smaller values of $x$ and larger values of $p$ are required to reach a net information rate of zero. This corresponds to the case where two mixed squeezed states are as equally likely to be distinguished as two pure squeezed states.

We also considered the practical case where a homodyne detection measurement is used to distinguish the pure and mixed coherent states and compared the probability of error in those situations to the POVM measurement of the trace distance. As expected, we find that the POVM outperforms the projective measurement of the homodyne detector, i.e., it reduces the probability of error. 

Acknowledgments --  We thank the support of the Australian Research Council (ARC) and discussions with Mile Gu, Nathan Walk and Andrew Lance. C.W. would like to thank Stefano Olivares, Gerd Leuchs and Christoffer Wittmann for pointing out additional references.


\begin{thebibliography}{99}

\bibitem{Nielsen00a} M.A. Nielsen and I.L. Chuang, {\it Quantum Computation and Quantum Information} (Cambridge University Press, Cambridge, 2000).

\bibitem{Helstrom76a} C. W. Helstrom, {\it Quantum Detection and Estimation Theory},
Mathematics in Science and Engineering, vol. 123 (Academic
Press, New York, 1976).

\bibitem{S.L.Braunstein2005} S. L. Braunstein and P. van Loock, Rev. Mod. Phys. \textbf{77}, 513
(2005).

\bibitem{Cerf2007} N. J. Cerf, G. Leuchs, and E. S. Polzik, eds., {\it Quantum Information
with Continuous Variables of Atoms and Light} (Imperial
College Press, 2007).

\bibitem{A.Chefles2000} A. Chefles, Contemp. Phys. \textbf{41}, 401 (2000).

\bibitem{Bergou2004} J. Bergou, U. Herzog, and M. Hillery, \textit{Discrimination of Quantum
States} (Springer-Verlag, 2004).

\bibitem{J.Twamley1996} J. Twamley, J. Phys. A: Math. Gen. \textbf{31}, 3659 (1996).

\bibitem{Gh.-S.Paraoanu1998} Gh.-S.Paraoanu and H. Scutaru, Phys. Rev. A \textbf{58}, 869 (1998).

\bibitem{A.Chefles1998} A. Chefles and S. M. Barnett, Phys. Lett. A \textbf{250}, 223 (1998).

\bibitem{Elk2002} S. van Elk, Phys. Rev. A \textbf{66}, 042313 (2002).

\bibitem{Oli04} S.~Olivares and M.~G.~A.~Paris, J.Opt.B: Quantum Semiclass. Opt. \textbf{6}, 69 (2004).

\bibitem{H.Nha2005} H. Nha and H. J. Carmichael, Phys. Rev. A \textbf{71}, 032336 (2005).

\bibitem{Andersen2006} U. L. Andersen, M. Sabuncu, R. Filip, and G. Leuchs, Phys.
Rev. Lett. \textbf{96}, 020409 (2006).

\bibitem{Calsamiglia2008} J. Calsamiglia, R. Munoz-Tapia, L. Masanes, A. Acin, and
E. Bagan, Phys. Rev. A \textbf{77}, 032311 (2008).

\bibitem{S.Pirandola2008b} S. Pirandola and S. Lloyd, Phys. Rev. A \textbf{78}, 012331 (2008).

\bibitem{Dol73} S.~Dolinar, Research Laboratory of Electronics, MIT, Quarterly Progress Report No. 111, 1973, p. 115.

\bibitem{Ken73} R.~S.~Kennedy, Research Laboratory of Electronics, MIT, Quarterly Progress Report No. 108, 1973, p. 219.

\bibitem{Tak05} M.~Takeoka \textit{et al.}, Phys. Rev. A \textbf{71}, 022318 (2005).

\bibitem{Sas96} M.~Sasaki \textit{et al.}, Phys. Rev. A \textbf{54}, 2728 (1996).

\bibitem{Tak08} M.~Takeoka \textit{et al.}, Phys. Rev. A \textbf{78}, 022320 (2008).

\bibitem{Ger04} J.~M.~Geremia, Phys. Rev. A \textbf{70}, 062303 (2004).

\bibitem{Coo07} R.~L.~Cook \textit{et al.}, Nature (London) \textbf{446}, 774 (2007).

\bibitem{Wit08} C.Wittmann \textit{et al.}, Phys Rev. Lett. \textbf{101}, 210501 (2008).

\bibitem{F.Grosshans2003} F. Grosshans, G. Assche, J. Wenger, R. Brouri, N. J. Cerf, and
P. Grangier, Nature \textbf{421}, 238 (2003).

\bibitem{F.Grosshans2002} F. Grosshans and P. Grangier, Phys. Rev. Lett. \textbf{88}, 057902
(2002).

\bibitem{Silberhorn2002} C. Silberhorn, T. C. Ralph, N. Lutkenhaus, and G. Leuchs,
Phys. Rev. Lett. \textbf{89}, 167901 (2002).

\bibitem{C.Weedbrook2004} C. Weedbrook, A. M. Lance, W. P. Bowen, T. Symul, T. C.
Ralph, and P. K. Lam, Phys. Rev. Lett. \textbf{93}, 170504 (2004).

\bibitem{S.Pirandola2008} S. Pirandola, S. Mancini, S. Lloyd, and S. L. Braunstein, Nature
Physics 4, \textbf{726} (2008).

\bibitem{M.Heid2007} M. Heid and N. Lutkenhaus, Phys. Rev. A \textbf{76}, 022313 (2007).

\bibitem{Lance2005} A. M. Lance, T. Symul, V. Sharma, C.Weedbrook, T. C. Ralph,
and P. K. Lam, Phys. Rev. Lett. \textbf{95}, 180503 (2005).

\bibitem{T.Symul2007} T. Symul, D. J. Alton, S. M. Assad, A. M. Lance, C. Weedbrook,
T. C. Ralph, and P. K. Lam, Phys. Rev. A \textbf{76}, 030303
(2007).

\bibitem{Fuchs99a} C. A. Fuchs and J. van de Graaf, IEEE Trans. Inf. Theory \textbf{45},
1216 (1999).

\bibitem{M.Reed1972} M. Reed and B. Simon, {\it Methods of Modern Mathematical
Physics - Part I: Functional Analysis} (Academic Press, 1972).

\bibitem{A.Gilchrist2005} A. Gilchrist, N. K. Langford, and M. A. Nielsen, Phys. Rev. A
\textbf{71}, 062310 (2005).

\bibitem{C.C.Gerry2005} C. C. Gerry and P. L. Knight, {\it Introductory Quantum Optics}
(Cambridge University Press, Cambridge, 2005).

\bibitem{Shannon48a} C. E. Shannon, Bell System Tech. J. \textbf{27}, 379 (1948).

\bibitem{U.Leonhardt1997} U. Leonhardt, {\it Measuring the Quantum State of Light} (Cambridge
University Press, 1997).

\bibitem{Levitin95a} L. B. Levitin, in {\it Quantum Communications and Measurement},
edited by V. P. Belavkin, O. Hirota, and R. L. Hudson (Plenum
Press, New York, 1995), pp. 439–448.

\end{thebibliography}
\end{document}